\documentstyle[12pt]{article}

 at 10truept

\def\ga{\gamma}
\def\de{\delta}
\def\ep{\epsilon}

\def\la{\lambda}

\def\ph{\phi}

\def\ch{\chi}
\def\ps{\psi}

\def\Ga{\Gamma}
\def\De{\Delta}

\def\cl{{\cal L}}
\def\fr#1#2{{{#1} \over {#2}}}
\def\prt{\partial}

\def\vev#1{\langle {#1}\rangle}

\def\frac#1#2{{\textstyle{{#1}\over {#2}}}}

\def\lsim{\mathrel{\rlap{\lower4pt\hbox{\hskip1pt$\sim$}}
    \raise1pt\hbox{$<$}}}
\def\gsim{\mathrel{\rlap{\lower4pt\hbox{\hskip1pt$\sim$}}
    \raise1pt\hbox{$>$}}}
\def\sqr#1#2{{\vcenter{\vbox{\hrule height.#2pt
         \hbox{\vrule width.#2pt height#1pt \kern#1pt
         \vrule width.#2pt}
         \hrule height.#2pt}}}}

\def\Re{\hbox{Re}\,}
\def\Im{\hbox{Im}\,}

\newcommand{\beq}{\begin{equation}}
\newcommand{\eeq}{\end{equation}}
\newcommand{\bea}{\begin{eqnarray}}
\newcommand{\eea}{\end{eqnarray}}
\newcommand{\rf}[1]{(\ref{#1})}

\renewenvironment{thebibliography}[1]
 { \rm
   \begin{list}{\arabic{enumi}.}
    {\usecounter{enumi} \setlength{\parsep}{0pt}
     \setlength{\itemsep}{3pt} \settowidth{\labelwidth}{#1.}
     \sloppy
    }}{\end{list}}

\begin{document}

\baselineskip=12pt
\begin{flushright}
{IUHET 337\\}
{July 1996\\}
\end{flushright}
\vglue 5.5 truecm

\begin{center}
{\bf CPT, STRINGS, AND NEUTRAL-MESON SYSTEMS%
\footnote{%\tenrm 
Presented at the Workshop on $K$ Physics,
Orsay, France, May-June 1996 }
\\}
\vglue 0.8cm
{V. Alan Kosteleck\'y\\} 
\bigskip
{\it Physics Department, Indiana University\\}
\medskip
{\it Bloomington, IN 47405, U.S.A.\\}
\end{center}

\vglue 3truecm

{\rightskip=3pc\leftskip=3pc\noindent
This talk provides an overview of a
string-based mechanism for spontaneous CPT violation.
A summary is given of theoretical developments.
The mechanism could generate CPT-violating contributions
to a four-dimensional low-energy effective theory
and thereby produce detectable consequences 
for candidate experiments with neutral-meson systems.

}

\newpage

\baselineskip=20pt

\noindent
{\bf 1. Introduction}
\vglue 0.4 cm 

Invariance under CPT,
which is the combined operation of 
charge conjugation C,
parity reversal P,
and time reversal T,
is a basic theoretical feature of 
local relativistic field theories of point particles
\cite{cpt1}-\cite{cpt6}.
It has been tested experimentally 
to great precision in a broad variety of settings
\cite{pdg}-\cite{lo}.
The combination of the powerful theoretical result about CPT 
in particle physics together with 
the existence of high-precision experiments 
means that CPT violation is well suited as a candidate signature 
for new, non-particle physics such as string theory
\cite{kp1,kp11}.

Since strings are extended objects,
they violate assumptions normally imposed
in theoretical proofs of CPT invariance.
In fact,
a stringy mechanism 
that can induce spontaneous Lorentz symmetry breaking
\cite{ks}
with accompanying CPT violation is known
\cite{kp1,kp2}.
These effects are briefly described in section 2 below.

If spontaneous CPT violation occurs in a realistic theory,
then its consequences for low-energy physics can be modeled.
It turns out that effects could appear in any 
of the neutral-meson systems,
with specific experimental signatures.
Various schemes can be examined for detecting CPT violation 
in each neutral-meson system.
Potentially interesting experiments can be performed
with either correlated or uncorrelated mesons.
Sections 3, 4 and 5 below contain a brief description 
of some aspects of the above ideas.
For more details about any of these topics,
the reader is referred to the original literature
on the $K$ system
\cite{kp1,kp11,kp3}, 
the two $B$ systems
\cite{kp3,ck1,kv},
and the $D$ system
\cite{kp3,ck2}. 

Note that the type of CPT violation discussed in
refs.\ \cite{kp1}-\cite{ck2}
is different from that possibly arising in the context
of quantum gravity
\cite{qg1}-\cite{qg3},
which involves unconventional quantum mechanics
and has an entirely different experimental signature
in the kaon system
\cite{qg4,qg5}.

\vglue 0.6 cm 
\noindent
{\bf 2. Spontaneous CPT Breaking}
\vglue 0.4 cm 

In any higher-dimensional 
Lorentz-invariant theory that purports to be realistic,
spontaneous Lorentz breaking
\it must \rm occur in some form.
In string theory, 
there exists a natural mechanism inducing spontaneous
Lorentz violation
\cite{ks}.
The origin of the effect is most readily seen at the
level of string field theory,
where it is due to the appearance of certain interactions 
that cannot occur in conventional four-dimensional 
renormalizable gauge theories.
The existence of these interactions is compatible
with string gauge invariance because string field theory
involves an infinite number of particle fields,
which in turn is a consequence of string nonlocality.

When nonzero expectation values for scalars arise,
these interactions can generate contributions 
to the mass matrices for Lorentz tensor fields. 
The signs of the contributions can be such as to 
trigger instabilities in the effective potentials
of the tensors,
which then in turn can acquire nonzero expectation values.
This causes spontaneous Lorentz breaking,
which can be accompanied by CPT breaking
\cite{kp1}.

The mechanism as described above is not specific to any model.
Some direct support for its occurrence in an explicit theory
emerges from the study of the solution space of 
the string field theory for the open bosonic string 
\cite{kp2}.
It is possible to construct analytically the leading
terms of the action in a level-truncation scheme.
The equations of motion can then be obtained,
and a search can be performed 
for Lorentz- and CPT-breaking extrema of the action 
that persist as the level number is increased
in the truncation scheme.
Over 20,000 nonvanishing terms in the action
have been examined in some cases.
Since a given solution takes the form 
of a set of expectation values,
their Lorentz- and CPT-breaking properties can be examined.
These are in agreement with features anticipated 
from the generic theoretical mechanism.

\vglue 0.6 cm 
\noindent
{\bf 3. Low-Energy Model}
\vglue 0.4 cm 

Assuming that the Universe can be modeled with
a realistic string theory,
the mechanism discussed in section 2
might produce CPT-violating contributions
to the effective four-dimensional low-energy theory,
which is the standard model. 
Since the Universe is found experimentally to be CPT invariant 
to a high degree of precision,
any CPT breaking must be suppressed.
A natural dimensionless suppression factor is provided by
the ratio $r \sim 10^{-17}$
of the low-energy scale to the Planck scale.
A suppression of this size can create effects 
of comparable or somewhat smaller
magnitude to the present experimental sensitivity
in the kaon and other neutral-meson systems 
\cite{kp1,kp3}.

To make this suggestion more definite,
one can consider a generic CPT-breaking contribution
to the effective low-energy theory
arising from a compactified string theory
\cite{kp11,kp3}:
\beq
\cl \sim \fr {\la} {M^k} 
\vev{T}\cdot\overline{\ps}\Ga(i\prt )^k\ch
+ {\textstyle h.c.}
\quad .
\label{a}
\eeq
Here, 
$\ps$ and $\ch$ are four-dimensional fermions
taken as component quark fields of the meson,
coupled via a gamma-matrix structure $\Ga$.
The expectation $\vev{T}$ involves a Lorentz tensor $T$.
The mass scale $M$ is an appropriate large scale
such as the Planck mass or a compactification scale,
and it compensates for the derivative couplings.
The parameter $\la$ is a dimensionless coupling constant.

\vglue 0.6 cm 
\noindent
{\bf 4. Neutral-Meson Systems}
\vglue 0.4 cm 

It is possible to analyse the effects of terms of the
form \rf{b}
on the effective hamiltonian governing the 
time evolution of a neutral-meson system.
At a purely phenomenological level,
CP violation in a neutral-meson effective hamiltonian
can be parametrized by two quantities.
One governs T violation
and is usually denoted $\ep_K$ for the kaon system.
The other controls CPT violation
and is denoted $\de_K$.
Similar parameters exist for the other neutral mesons,
denoted generically as $P$ mesons in what follows.

The standard model provides a theoretical framework
for understanding the origin of a nonzero value of $\ep_P$
via the CKM matrix.
Analogously,
string-inspired spontaneous CPT breaking 
of the form \rf{a}
in the low-energy effective theory
could be used to understand a nonzero value of $\de_P$. 
It can be shown that
\cite{kp11,kp3}
\beq
\de_P = i 
\fr{h_{q_1} - h_{q_2}}
{\sqrt{\De m^2 + \De \ga ^2/4}}
e^{i\hat\ph}
\quad ,
\label{b}
\eeq
where the factors $h_{q_j}=r_{q_j}\la_{q_j}\vev{T}$
arise from the coefficients of the interaction \rf{a}
and from contributions $r_{q_j}$ 
from the quark-gluon sea in the meson.
The mass and rate differences $\De m$ and $\De\ga$
are experimental observables,
as is the phase $\hat\ph = \tan^{-1}(2\De m/\De \ga)$.
It follows that the string scenario predicts
the relationship
\beq
\Im \de_P = \pm \cot\hat\ph ~\Re\de_P
\quad .
\label{c}
\eeq
It can also be shown that contributions
from this mechanism 
to direct CPT violation in decay amplitudes 
are negligible.

\vglue 0.6 cm
\noindent
{\bf 5. Experimental Tests in Neutral-Meson Systems}
\vglue 0.4 cm

There are several neutral-meson systems that can be
considered as candidates for CPT tests:
the $K$, $D$, $B_d$, and $B_s$ systems
\cite{kp3}.
In this regard,
note one especially interesting feature 
of the theoretical expression
for the CPT-violating parameter $\de_P$
in the string scenario:
its magnitude depends on couplings 
that could differ for distinct quark flavors,
as for example do the Yukawa couplings 
in the standard model.
It is therefore of interest to test for CPT breaking 
in neutral-meson systems other than the $K$ system.
As an extreme example, 
it is presently even possible that CPT violation
in the $B$ system could be \it larger \rm than 
the expected (but as yet unobserved) conventional T violation,
even though the corresponding possibility
is experimentally excluded for the kaon system.

In each neutral-meson system,
experiments can be envisaged that involve
either correlated mesons 
produced via decay of appropriate quarkonia
or uncorrelated ones produced by a variety of means.
At present,
a basic theoretical catalogue of interesting experimental
asymmetries has been created for all the systems.
For some cases,
the theoretical analysis now includes Monte-Carlo simulations 
of realistic experimental data incorporating acceptances
and background effects.
Here are a few very brief comments
about results for the $K$, $D$, and $B$ systems in turn.
More details can be found in the original literature
\cite{kp1}-\cite{ck2}.

At present,
CPT violation in the neutral-meson systems is bounded
only in the kaon case.
The current limits on $\Re\de_K$ 
are of order $10^{-3}$
while those on $\Im\de_K$
are of order $10^{-4}$
\cite{expt1}-\cite{expt5}.
Under favorable circumstances,
perhaps incorporating data from $\ph$ factories
\cite{ph1}-\cite{ph3},
reliable bounds improved by an order of magnitude
could be envisaged.

Attainable bounds on CPT using the $D$ system are harder
to estimate because $D$ mixing has not been observed
and has theoretically uncertain magnitude.
However, 
possibilities for placing  
definite experimental bounds on CPT violation 
could include limiting direct CPT violation
and bounding one or both of the $K$- and $D$-system parameters
$\de_K$ and $\de_D$.

There is presently no limit in the literature on 
the CPT-violating parameter $\de_B$ for the $B_d$ system.
This is an especially interesting arena for future exploration
because the $b$ quark is involved
and in certain string scenarios CPT breaking 
could be greatest there.
It has recently been shown that 
sufficient data have already been taken 
to limit the CPT-violating parameter $\de_B$ at about
the 10\% level
\cite{kv}.
This would be comparable to the current bound on conventional
indirect T violation in the $B$ system,
which is presently of order 5\%.
Tighter limits could be attained in planned experiments
at various $B$ factories.

\vglue 0.6 cm
{\bf\noindent Acknowledgments}
\vglue 0.4 cm

I thank Don Colladay, Rob Potting, Stuart Samuel, 
and Rick Van Kooten for enjoyable collaborations 
on the topic of this talk.
This work was supported in part
by the United States Department of Energy 
under grant number DE-FG02-91ER40661.

\vglue 0.6 cm
{\bf\noindent References}
\vglue 0.4 cm


\begin{thebibliography}{xx}
\bibitem{gg}
A summary of non-$K$ tests 
is given by G. Gabrielse, these proceedings.
\end{thebibliography}
\end{document}